\documentclass{pasj00}

\SetRunningHead{Astronomical Society of Japan}{Usage of \texttt{pasj00.cls}}

\title{Threshold between spontaneous and cloud-collisional star
formation}

\author{Shinya \textsc{Komugi}, Yoshiaki \textsc{Sofue} \and Fumi \textsc{Egusa}}
\affil{Institute of Astronomy, The University of Tokyo, 2--21--1 Osawa,
Mitaka-shi, Tokyo, 181--8588}
\email{skomugi@ioa.s.u-tokyo.ac.jp}


\Received{$\langle$reception date$\rangle$}
\Accepted{$\langle$acceptance date$\rangle$}
\Published{$\langle$publication date$\rangle$}
\begin{document}
\maketitle

\begin{abstract}
Based on simple physical and
 geometric assumptions, we have calculated the mean surface molecular
 density of spiral galaxies 
 at the threshold between star formation
 induced by cloud-cloud collision and spontaneous gravitational
 collapse.  The calculated threshold is approximately  $\log \Sigma_\mathrm{crit} \sim 2.5$,
 where $ \Sigma \quad \mathrm{M_{\solar}}\cdot \mathrm{pc}^{-2}$ is the observed surface mass density of an assumed flat
 gas disk.  Above this
 limit, the rate of molecular cloud collisions dominates over spontaneous molecular
 cloud collapse.  This model may explain the apparent discontinuity in the
 Schmidt law found recently at $2 \lesssim \log \Sigma \lesssim 3$.
 \end{abstract}

\section{Spontaneous and collisional star formation}
The onset of star formation within an ensemble of molecular clouds is
thought to have two modes, spontaneous (stochastic)
gravitational collapse \citep{bonnell04, krumholtz05}, and
collisions between molecular clouds (e.g., \cite{tan}).  Most
studies to date have focused on one of these two possibilities.
In actual physical situations which cover a range of physical
parameters, these
two processes should both have effect on the overall star formation in
galaxy disks.  In regions of higher molecular density, collisions should
become dominant over spontaneous formation, because collisions will
occur on timescales shorter than spontaneous star formation.  In regions
with scarse cloud number density, collisions should also be rare and
therefore star formation from other stochastic means should dominate.  The
transition from spontaneous to collisional star formation may have
observed effects on the relation between molecular gas content and star
formation rate (SFR).  This expected transition has not been clearly
observed yet, however.

 There is growing observational evidence that cloud collision can trigger star
 formation \citep{loren76, scoville86, hasegawa94, koda06}.  These
 studies show that cloud collisions are seen mainly at high density
 regions, and therefore
 merging and interacting galaxies are candidates for galaxies forming
 stars mainly by collision. A comparison of these galaxies with 
normal galaxies have been conducted by \citet{young86}.  They have
found that merging/interacting galaxies exhibit SFRs
that are an order of magnitude higher than normal galaxies of the same molecular gas
density.  Similarly, central regions and spiral arms of normal
galaxies, which generally have high molecular densities, may include
star formation induced by collision.  However, quantitative calculations which
give direct (observable) parameters on exactly $\textit{where}$ in
parameter space star formation
transits from spontaneous to collisional mechanisms have not been
proposed.     
\section{Model}
\begin{figure}[t]
  \begin{center}
    \FigureFile(80mm,80mm){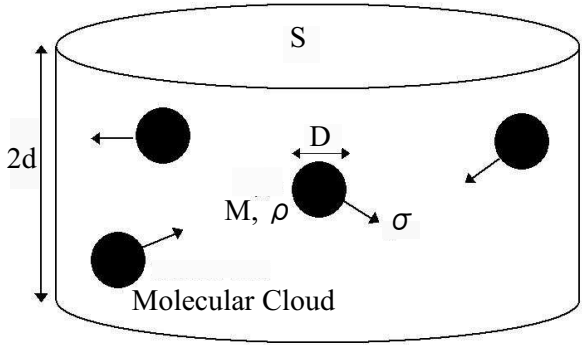}
  \end{center}
  \caption{Schematic picture of model.  The base of the cylinder is 
 parallel to the galactic disk.  Each of the molecular clouds have size
 $D$, mass $M$, density $\rho$, and velocity $\sigma$.}\label{fig1}
\end{figure}

We consider a spiral galaxy that can be modelled as a thin disk.  We
take a cylinder of height $2d$ and base area $S$ within the same plane
as the galaxy disk (see schematic figure \ref{fig1}).  We can think of $d$ as the scale
height, and $S$ as the area of the deprojected beam of the observation.  Within this
cyclinder are $n$ spherical molecular clouds with diameter $D$, mass $M$ and mean
molecular density $\rho$.  Assuming that the observer can see through the disk and
derive a surface molecular gas density $\Sigma$, the obvious equations are:
\begin{equation}
\Sigma=\frac{nM}{S}
\end{equation}
\begin{equation}
M=\frac{4\pi}{3}\left(\frac{D}{2}\right)^3 \rho
\end{equation}

These molecular clouds are moving about the cylinder with an inter-cloud velocity
dispersion $\sigma$.

For further analysis, we have several simple assumptions:
\begin{itemize}

\item
star formation occurs on the free fall timescale, dependent only on the
     initial density
\item
molecular clouds are moving about the thin cylinder in random motion

\item
star formation in galactic disks is a two dimensional process.
\end{itemize}
 
The third assumption is motivated from the fact that giant molecular
clouds have size comparable to the scale height $d$ \citep{stark, malhotra}.  We can therefore assume that all processes occur in
two dimensions and we can hereafter ignore $d$ hereafter.

We have
\begin{equation}
t_\mathrm{ff}=\left(\frac{3\pi}{32\mathrm{G}\rho}\right)^{\frac{1}{2}}
\end{equation}
where $t_\mathrm{ff}$ is the free fall timescale and G the gravitational
constant.  Collisions between molecular clouds should have a mean free path
$\lambda_\mathrm{mfp}$, for two dimensions, given by
\begin{equation}
\lambda_\mathrm{mfp}=\frac{1}{\sqrt{2}ND}
\end{equation}
where $N=n/S$ is the number density of molecular clouds.
A GMC harboring spontaneous star formation will move a linear distance
of $\sigma t_\mathrm{ff}$ during its formation of stars.  In a regime
where spontaneous star formation is dominant over collision, parameters
should satisfy the relation
\[
 \sigma t_\mathrm{ff} \ll \lambda_\mathrm{mfp}
\]  
so that the GMC does not collide with other GMCs during formation.  On
the other hand, a regime where collision is dominant should satisfy
\[
 \sigma t_\mathrm{ff} \gg \lambda_\mathrm{mfp}.
\]
Therefore, if a threshold exists between these two regimes, parameters
should satisfy the equation
\begin{equation}
 \sigma t_\mathrm{ff} \approx \lambda_\mathrm{mfp}.
\end{equation}
Substituting equations (1) through (4) into (5) and denoting the critical
$\Sigma$ as $\Sigma_\mathrm{crit}$, we obtain

\[
 \log \frac{\Sigma_\mathrm{crit}}{\mathrm{[M_{\solar}]}}= 
\]
\begin{equation}
-1.35+2\log \frac{D}{\mathrm{[pc]}} + 1.5\log \frac{\rho}{\mathrm{[M_{\solar}}
\cdot  \mathrm{pc}^{-3}]} - 
\log \frac{\sigma}{[\mathrm{km} \cdot \mathrm{s}^{-1}]}
\end{equation}

\section{$\Sigma_\mathrm{crit}$ value}
Now we apply actual values to equation (6).  For typical Galactic molecular
clouds with $D \sim 60 [\mathrm{pc}]$ and $\rho \sim 30
[\mathrm{cm}^{-3}] = 1.5 [\mathrm{M}_{\solar} \cdot \mathrm{pc}^{-3}]$,
and inter-cloud velocity dispersion $\sigma \sim 4 [\mathrm{km} \cdot \mathrm{s}^{-1}]$,
we obtain $\log \Sigma_\mathrm{crit}=  1.91 \quad
[\mathrm{M_{\solar}} \mathrm{pc}^{-2}]$.  For observed molecular surface
densities above this value, collisions dominate.   

We can further reduce the number of variables.  \citet{dame86}
have found that for Galactic molecular clouds, $\rho$ scales with $D$
such that $\log \rho = 2.80 - 1.32 \log D $.  Substituting this relation
into equation (6) gives
\begin{equation}
\log \frac{\Sigma_\mathrm{crit}}{\mathrm{[M_{\solar}]}}=2.85+0.02\log \frac{D}{ [\mathrm{pc}]} -\log \frac{\sigma}{ [\mathrm{km}\cdot \mathrm{s}^{-1}]}
\end{equation}
It is apparent from equation (7), that the critical density is
practically independent of GMC diameter.  The critical surface density
is dependent mainly on cloud velocity dispersion $\sigma$, which is
generally taken to be below 10 $\mathrm{km}\cdot \mathrm{s}^{-1}$.
Table 1 lists values of $\Sigma_\mathrm{crit}$ for a range of
parameters to evaluate the lowest and highest cases.

Errors for this value can be evaluated based on the uncertainty of the
parameters.  $\sigma$ derived from other studies, numerical and
observational, have agreed-upon values between 1 [km/s] and 10[km/s].  The
value of $D$, which must be taken so that $D$ is representative of the
size which dominates star formation, is between 10 [pc] and
several$\times 100$[pc], although uncertainties in $D$ have little effect
on $\Sigma_\mathrm{crit}$.  On the average, we may safely apply
errors so that $\Sigma_\mathrm{crit}=2.8^{+0.1}_{-1.0}$

\begin{table}
  \caption{Derived values of $\Sigma_\mathrm{crit}$}\label{data}
  \begin{center}
    \begin{tabular}{ccc}  \hline \hline
     $D$ [pc]  & $\sigma$ [km$\mathrm{s}^{-1}$]& $\log \Sigma_\mathrm{crit}$ [$\mathrm{M_{\solar}}\mathrm{pc}^{-2}$] \\ \hline
      10 & 1  &  2.87 \\
         & 5  &  2.17 \\
         & 10 &  1.87 \\
     100 & 1  &  2.89 \\
         & 5  &  2.19 \\
         & 10 &  1.89  \\ \hline  
     \end{tabular}
    \end{center}
        Notes- $D$, the diameter of molecular clouds, are taken so that
 it may represent smallest (10pc) and largest (300pc) clouds.  $\sigma$
 from literature ranges from 4 km/s \citep{liszt81, clemens85} to 7km/s (\cite{stark89}), but we take values from 1
 to 10 km/s to demonstrate the range of values $\Sigma_\mathrm{crit}$
 may take. 
 \end{table}

\section{Comparison with observation}
The empirical relation between gas density and SFR, often referred to as
the Schmidt law, has long been
studied, and the relation of the form $\Sigma_\mathrm{SFR} \propto
\Sigma^N$, where $\Sigma_\mathrm{SFR}$ is the area averaged
SFR, has been extensively studied.  Details can be found elsewhere
(e.g., \cite{buat89}, \cite{buat92}, \cite{K98},
\cite{Rownd99},\cite{wong02}, \cite{Boissier}, \cite{Yao},
\cite{Gao04}, \cite{Heyer}).
This law, however, owes its dynamic range to the contrast between starburst galaxies and normal spiral galaxies.
  The starbursts have higher molecular density whereas normal spirals have considerably lower molecular density, so 
the combination of these two types of galaxies contributes dominantly to the linear fit of the Schmidt type power law. 
  
\begin{figure}
  \begin{center}
    \FigureFile(100mm,100mm){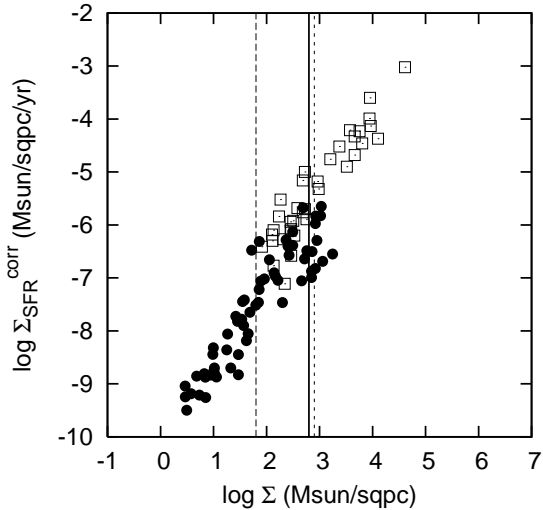}
  \end{center}
  \caption{The Schmidt law, from \citet{komugi}.  Abscissa is $\log \Sigma \quad \mathrm{M_{\solar}}\cdot \mathrm{pc}^{-2}$ 
and ordinate $\log \Sigma_\mathrm{SFR} \quad \mathrm{M_{\solar}}\cdot \mathrm{yr}^{-1} \mathrm{pc}^{-2}$.  
Filled circles represent normal galaxies, and open squares represent circumnuclear starbursts.  
The center vertical line is the value for $\Sigma_\mathrm{crit}=2.8$, where the dashed
 lines represent errors as stated in text.}\label{fig2}
\end{figure}

Any systematic differences in the Schmidt law between these two types of galaxies are ignored by fitting both with the same function.
Indeed, a possible significant difference in these two types
of galaxies were found by \citet{komugi}, by the study of high
molecular density central regions of normal spirals.  The Schmidt law,
with normal galaxies and circumnuclear starbursts superposed, is shown
in figure \ref{fig2}. 

An apparent discontinuity exists in the two sequences, and the critical surface density
 $\Sigma_\mathrm{crit}=2.8^{+0.1}_{-1.0}$, shown as a vertical line, coincides with the
transition.  Assuming that this apparent transition is real, it may be
explained by change in the physical nature of star formation, namely the
transition from spontaneous to cloud-cloud collisional star formation,
induced by the increase in molecular density.  

It must be pointed out that the SFR for the circumnuclear starbursts
are derived from far-infrared (FIR) luminosities, and those of normal galaxies
from internal extinction corrected H$\alpha$ luminosities (see
\citet{komugi} for details).  

These two methods of deriving the SFR are known to differ, in the sense that FIR gives systematically higher SFRs
 compared to H$\alpha$.  FIR derived SFRs overestimate the \textit{true} current SFR because it is contaminated with 
emission from dust heated by lower mass stars.  H$\alpha$ emission is subject to interstellar dust extinction, therefore
underestimating the current SFR.  These may lead the reader to believe
 that the discontinuity is not a \textit{true} one due to physical
 processes.  In figure 2, however, the SFR for normal galaxies is derived from internal extinction corrected
H$\alpha$ luminosity, found by \citet{kewley} to agree with FIR based SFR within 10\%.  The order-of-magnitude offset between the 
two types of galaxies therefore can not be attributed to such observational shorthands.  The unaccountable offset may be 
due to differences in the physical process of star formation.  \citet{elmegreen89} has suggested that a factor 4 decrease in 
$t_\mathrm{ff}$ will occur at a shocked surface of a molecular cloud, and this can account for much of the rise in SFR for starbursts,
which lies in the "collision" regime in figure 2.

\section{Discussion}
\subsection{Other parameters}
The simplicity of equation (5) poses a question of whether this analysis
is not an oversimplification.  Therefore, we go on to consider other
parameters which may modify equation (5).  The basic assumptions made in section 1 is that 
we have assumed that star formation occurs on the free
fall timescale $t_\mathrm{ff}$, but the timescale of star formation is
still an unresolved issue, and studies often give longer timescales than
$t_\mathrm{ff}$.
Second, we have assumed a random motion of the
molecular clouds based on epicyclic motion of the clouds, while in reality, differential rotation in
galactic disks is also important to the overall motion of
clouds.

The star formation timescale $t_\mathrm{sf}$, is equal to
$t_\mathrm{ff} \sim 10^6 [\mathrm{yr}]$ when star formation is dominated
by gravitational pertubations, and this type of formulation
$t_\mathrm{sf} \sim t_\mathrm{ff}$ has been adopted in many studies.
Other studies argue that $t_\mathrm{sf}$ is longer than $t_\mathrm{ff}$
because magnetic fields provide support against gravitational
contraction, and star formation is impossible without ambipolar
diffusion \citep{tan} in which case the star formation timescale is
 longer by an order of magnitude \citep{McKee}.  Still, some observational studies suggest that star
formation occurs on scales of $\sim 10^6 [\mathrm{yr}]$, based on
the observed offset between CO and H$\alpha$ arm in spirals \citep{egusa}.
  This issue is not resolved, and we adopt $t_\mathrm{sf} \sim t_\mathrm{ff}$ for
simplicity.  However, if $t_\mathrm{sf} > t_\mathrm{ff}$ by an order or
more, collisional star formation will dominate at a much lower observed
density of about $\Sigma_\mathrm{crit}=1.8-\log \sigma$.
The differential rotation of galactic disks causes molecular clouds to
collide as in the case of random motion.  \citet{gammie} find
$\sigma = 5.1 \quad [\mathrm{km \cdot s^{-1}}]$ for cloud encounters due
differential rotation, whereas random velocity dispersion of GMCs are
$\sim 7 \quad[\mathrm{km \cdot s^{-1}}]$(See \cite{stark89}), or $\sim
4 \quad [\mathrm{km \cdot s^{-1}}]$ \citep{liszt81, clemens85}.
Therefore, we can say that the value of cloud collision velocity,
$\sigma$, is comparable even with the assumption of differential
rotation.

\subsection{Mergers and interacting galaxies}
The argument of threshold surface density proposed in this Letter cannot
specify what kind of change will appear in the
$\Sigma_\mathrm{SFR}-\Sigma$ relation such as that seen in
figure 2.  The discontinuity may well be a result of transition from
spontaneous to collisional star formation, but our results do not
predict or necessitate a discontinuity.  A sudden rise in SFR
as collision dominates, however, is physically intriguing and easy to understand.  In a
collision between two GMCs, the shocked region between the clouds are
compressed so that the density abruptly becomes higher in this layer compared
to other parts of the GMC.  This will shorten $t_\mathrm{ff}$, and
consequently the SFR is higher (\citet{elmegreen89} finds that
$t_\mathrm{ff}$ is shortened by a factor of 0.25): this should appear as a discontinuity in
the Schmidt law.  

In this respect, it is reasonable to expect that merging and interacting
galaxies form stars mainly by cloud-cloud collision,
and therefore exhibit higher SFR than galaxies with the same molecular
gas density.  In these systems, the critical density $\Sigma_\mathrm{crit}$ is far lower than normal spirals, because GMCs in colliding
galaxies will have far higher collision velocity $\sigma$.  In a collision between two
spirals, the typical rotation velocity of the galaxies will become
representative of $\sigma$, with about $\sim 200 [\mathrm{km \cdot
s^{-1}}]$, depending on the sense of rotation of the colliding spirals.
In this case, $\log \Sigma_\mathrm{crit}$ can be 0 to 1, the
typical global surface molecular mass density of spirals, so
that most of the star formation can be attributed to cloud-cloud
collision.

\section{Conclusion}
We have constructed a simple model to consider the threshold between
star formation dominated by spontaneous gravitational collapse and
cloud-cloud collision.  Although both cloud-cloud collision and gravitational collapse processes
have for decades been thought to cause star formation, the threshold of these two
processes has not attracted much attention.  Most
studies of collisional star formation have focused on
qualitative aspects.  Numerical studies such as that by \citet{mazzei} have pointed out a threshold, 
but with the contradictory result
that collisions between clouds result in lower star formation rates.  

Our model, by assuming that star formation is a bimodal process
dominated either by spontaneous gravitational collapse or cloud-cloud
collision, does not take into account the possiblity of star formation
induced by expanding SNR shells (self-propagating star formation), or
complicated structures in temperatures or chemistry.  Simple as our model
is, it succeeds in explaining the apparent discontinuity in the Schmidt law.
Moreoever, our formulation coupled with the empirical implication that most
clouds are virialized and thus GMC size and mean density are
anticorrelated, leads to the conclusion that the threshold surface mass
density $\log \Sigma_\mathrm{crit}$ is practically independent of cloud parameters (size, density,
mass), but that it depends on the mean velocity dispersion $\sigma$ of
clouds.

Further tests of our formulation can be expected from observations of
mergers.  In merging spiral pairs whose
rotation are oriented in the reverse direction, the relative velocity of
the clouds are increased so that according to equation (7), collisional
star formation starts at a very low surface mass density.  However, if
the rotation is oriented in the same direction, $\sigma$ will not
increase as much as in the reverse orientation.  Thus, the threshold
$\log \Sigma_\mathrm{crit}$
starts at a higher density.  Assuming that collisional star formation
results in a discontinuity in the SFR relation with surface gas density,
we can compare mergers with these two
orientations with average normal galaxies which we can assume to be
dominated by a spontaneous process.  Then, we should see that mergers
with reverse rotation orientation will start its discontinuity in the
Schmidt law at a lower surface mass density than mergers with ordered rotation.\\[0.5cm]

The authors thank H. Nakanishi, and S. Onodera, and K. Kohno for fruitful discussions.  S.K. and F.E. were
financially
supported by a Research Fellowship from the Japan Society for the Promotion
of Science for Young Scientists.

\end{document}